\documentclass[twocolumn,aps,pra,showpacs]{revtex4}
\usepackage{amssymb}
\usepackage{graphics}
\newcommand{\be}{\begin{equation}}
\newcommand{\ee}{\end{equation}}
\newcommand{\bea}{\begin{eqnarray}}
\newcommand{\eea}{\end{eqnarray}}

\begin{document}

\title{Gravitational self-localization in quantum measurement}
\author{Tam\'as Geszti}
\affiliation{Department of the Physics of Complex Systems, \\
E{\"o}tv{\"o}s University; H-1117 Budapest, Hungary \\
{e-mail: \tt geszti@galahad.elte.hu}
}

\begin{abstract}
Within Newton-Schr{\"o}dinger quantum mechanics which allows gravitational
self-interaction, it is shown that a no-split no-collapse measurement
scenario is possible. A macroscopic pointer moves at low acceleration,
controlled by the Ehrenfest-averaged force acting on it. That makes
classicality self-sustaining, resolves Everett's paradox, and outlines
a way to spontaneous emergence of quantum randomness. Numerical
estimates indicate that enhanced short-range gravitational forces are
needed for the scenario to work. The scheme fails to explain quantum 
nonlocality, including two-detector anticorrelations, which points 
towards the need of a nonlocal modification of the Newton-Schr{\"o}dinger 
coupling scheme.
\end{abstract}

\maketitle

\section{INTRODUCTION}
If microscopic quantum mechanics and macroscopic classical mechanics
are limiting cases of the same dynamic law, that law cannot be the
linear, deterministic Schr{\"o}dinger equation: the quantum measurement
process, with its random switch governed by squared amplitudes is manifestly
nonlinear, and apparently non-deterministic \cite{whz}. However, in
all the rest of quantum phenomena, nonlinearity is extremely elusive.
The concept of collapse (reduction) of the wave function \cite{n},
the applicability of which is limited to the act of quantum
measurement, is a proper expression of that situation.

As an attempt to find a unified dynamical law interpolating between
quantum and classical realms, modified quantum mechanics schemes of 
the so-called stochastic reduction type have been developed \cite{grwd}, 
which add both a nonlinearity and a random force to standard quantum
mechanics; those schemes are sometimes related to gravity \cite{qgr}.
Along with that line of thinking, there are continuing attempts to 
understand the effects of nonlinearity alone \cite{bbmw}, without 
built-in randomness. Besides calculating eventual detectable 
modifications of spectra, such studies are usually focussed on 
demonstrating the stabilization of soliton-like motions against the 
spreading of wave packets predicted by linear quantum mechanics: that 
stabilization is tentatively identified with emerging classicality. 
Nonlinearities which are too weak to show up mechanically, might prove 
strong enough to be detected by interferometry \cite{shinl}.

In what follows, we remain with the special case of an attractive 
nonlinearity provided by gravity, acting in a mean-field way\cite{di}: 
dynamics is governed by Schr{\"o}dinger's equation, in which gravity
appears as an external, classical field $\Phi$, obeying the Poisson
equation
\be
\nabla^2\Phi(\vec r) = 4\pi G\varrho(\vec r),
\label{poisson}
\ee
where $G$ is Newton's constant of gravity. The source term is the
quantum mechanical mean value of the operator of mass density in
quantum state $|\Psi\rangle$for particles of masses $m_i$ and position
operators $\hat{\vec r_i}$: 
\be
\varrho(\vec r) = 
\langle\Psi|\sum_i m_i\delta(\vec r-\hat{\vec r_i})|\Psi\rangle.
\label{source}
\ee
The $|\Psi\rangle$-dependence of the field $\Phi$ makes the
Schr{\"o}dinger equation nonlinear. In recent times, such a
scheme is called {\em Newton-Schr{\"o}dinger quantum mechanics,} and
considered a promising starting point to attack the problem of quantum
state reduction. 

In the present paper we pursue that line of reasoning, and focus our
attention to the situation of quantum measurement on a superposition,
which is a special situation in that the apparatus is actively driven 
towards different positions. That poses much more stringent
requirements on the attractive nonlinearity than the stabilization 
of a single wave packet does. Gravity-based nonlinearity \cite{di} is 
far more efficient in such a situation than any local one \cite{bbmw}, 
because of the long-range attraction between split components of a wave
packet. A further crucial point is that the strength of gravity is 
tested against acceleration; in quantum measurement microscopic bodies 
act on macroscopic ones, and the acceleration of the latter is low.

Di{\'o}si \cite{di} poses the question whether self-gravity excludes
persistently split wave packets, and answers no. Here we pose a
different question: can self-gravity exclude the {\em creation} of 
split wave packets under some circumstances; in particular,  can it 
successfully protect macroscopic bodies against macroscopic splitting, 
even in quantum measurement situations? As shown below, now the answer 
can be yes, if the bodies are massive enough and the accelerations are 
low. The detailed condition is derived below, see Eq. (\ref{class}).

If that condition is satisfied by macroscopic bodies, the
consequences are far-reaching and beneficial for the theory.
Classicality, once generated \cite{inf}, may become self-sustaining.
The nature of the quantum-classical border - the Bohrean cut - is 
immediately outlined: gravitational self-interaction is weak on 
microscopic objects, strong on macroscopic ones, and the two can be
distinguished by means of a physical criterion. Having a single 
dynamical law, Everett's no-collapse scheme \cite{ev} is recovered, 
without its paradoxical features: for a macroscopic body (e.g. a
detector) the split components of the wave function remain glued to 
each other, in accordance with everyday experience. Viewed from the
macroscopic side, this is effectively a {\em no-split no-collapse} 
scheme, in which Schr{\"o}dinger's cat remains doubly-dead or
doubly-live. The Schr{\"o}dinger wave field gets ``materialized'', 
being the source of gravity, thereby eliminating the necessity of 
philosophical discussions about epistemological vs. ontological 
status of the state vector. In addition, the scheme is suggesting 
dynamical mechanisms about generation of randomness in the emerging 
classical domain, in accordance with Born's rule of probabilities 
\cite{bn}.

In spite of that attractive view opening up, there are important
reasons to doubt in the validity of the Newton-Schr{\"o}dinger
scheme. First is the weakness of gravity as we know it from
macroscopic measurements: the numerical estimates presented below 
require a one-millimeter size of atomic condensed matter, and 
astronomical sizes of electron liquid to behave classically in a 
quantum measurement situation. If gravity proved to be strongly
enhanced with respect to Newton's law for short distances, the chances 
of the scheme would be much better; theoretical indications that such 
short-range enhancement may be real are briefly summarized in the text
below.

There is a more serious limitation of principle too: all of the
analysis in the next Section refers to the case of a single detector. 
As soon as several detectors are probing an entangled system \cite{epr} 
or only but a single microobject \cite{ei}, the scheme fails to describe
the much-tested quantum correlations manifest in multi-detector
anticoincidences, losing even the formal explanation offered by linear
quantum mechanics.

Below we will conclude that the tender spot of the scheme, responsible
for those important failures, is the way we characterize the source of
gravity. The obvious choice of Eq. (\ref{source}), used in all the work
listed under Ref. [\onlinecite{di}], may turn out to be wrong. Some
hints about the desired modification will be discussed at the end of
this paper. 

The main text below begins by deriving an effective
gravitational self-interaction potential that appears in the
c.o.m. Schr{\"o}dinger equation of a macroscopic body. The derivation
concludes in formulating the classicality condition, Eq. 
(\ref{class}). Then we continue by analyzing a minimum model of 
the measurement process, and demonstrate the way module-squared 
amplitudes control the macroscopic dynamics, offering a dynamical 
route to Born's rule. Then numerical estimates are presented, 
indicating that standard gravity is just strong enough to make the 
above picture valid for macroscopic condensed matter, however, it
is far too weak for all-electronic particle detection systems. The
analysis of entangled detectors and an outlook section conclude the
paper. 

\section{THE GRAVITATIONAL SELF-INTERACTION OF MACROSCOPIC BODIES}

Following Di{\'o}si [\onlinecite{di}], the detector is depicted as a
rigid body of mass density $\varrho_0$ and global mass $M$, the 
quantum state of which  can be separated into a frozen internal 
state and a c.o.m. wave function $\psi(\vec r,t)$. The latter obeys 
a c.o.m. Schr{\"o}dinger equation containing the potential energy 
$V_{gr}(\vec r)$ of the body, generated by itself at some place 
$\vec r~'$, averaged over $|\psi(\vec r~',t)|^2$ according to Eq. 
(\ref{source}). Using Di{\'o}si's result for the non-averaged interaction 
of two almost overlapping spheres of radius $R$, as long as spreading
and splitting of the c.o.m. wave packet are small: $|\vec r-\vec r~'|
\ll R$, we obtain the simple quadratic expression
\bea
V_{gr}(\vec r)\approx\int d\vec r~'\Bigl( V^0_{gr}
  +\frac12 M\omega_{gr}^2|\vec r-\vec r~'|^2\Bigr)
  |\psi(\vec r~',t)|^2\nonumber\\
 = V^0_{gr} + \frac12 M\omega_{gr}^2|\vec r-\vec r~^*|^2
   + \frac12 M\omega_{gr}^2~\Delta^2,
\label{gravpot}
\eea
where $\vec r~^*$ and $\Delta~(\ll R)$ are respectively the mean position and
the standard deviation of $|\psi(\vec r)|^2$,  
\be
V^0_{gr}=-\frac{6}{5}\frac{GM^2}{R},
\label{depth}
\ee
and
\be
\omega_{gr}^2 = G\varrho_0.
\label{gravfreq}
\ee
The analogous expressions for axial displacements in a slab-like geometry
are presented in the Appendix.

The last term in Eq. (\ref{gravpot}) is a time-dependent energy shift,
giving rise to a global phase factor, observable in principle by
interferometry as a scalar Aharonov-Bohm effect \cite{sab}.

The c.o.m. wave packet $\psi(\vec r,t)$, split or not, is expected to 
carry out all kinds of oscillatory motion in the harmonic potential 
(\ref{gravpot}). As a rough estimate, $\omega_{gr}^2 R$ is the maximum 
confining acceleration provided by gravitational self-attraction: 
that gives the classicality criterion
\be
\omega_{gr}^2 R \gg a_{max},
\label{class}
\ee
where $a_{max}$ is the maximum deconfining acceleration the apparatus 
undergoes during the measurement process. In view of Eq. (\ref{gravfreq}), 
this can be regarded as a criterion for the minimum size of the
apparatus to behave classically. Having a non-relativistic description 
at hand, the Planck scale is irrelevant in this respect; however, as 
seen below, the density of condensed matter points to a well-defined 
size range.

Eq. (\ref{class}) is subjected to numerical estimates below. However,
before that, we assume that the criterion is met and analyze its
consequences concerning quantum measurement.

\section{MODEL OF THE MEASUREMENT PROCESS} 

\subsection{Emergence of the averaged force}

As a minimum model of the compound system object+apparatus, we
investigate the measurement of the quasi-spin observable
$\hat\sigma_z$ on a two-state microobject with orthogonal basis
states $|+\rangle$ and $|-\rangle$. Throughout the measurement,
the state $|\Psi\rangle$ of the compound system is represented by the 
quasi-linear expression
\bea
\langle\vec r~|\Psi\rangle=\bigl(c_+|+\rangle
        +c_-|-\rangle\bigr)\psi(\vec r,0)~~~~~~~~~\nonumber\\
        ~~\Rightarrow c_+|+\rangle \psi_+(\vec r,t)
         +c_-|-\rangle\psi_-(\vec r,t).
\label{neumann}
\eea
Nonlinearity enters through the dynamics of the split partial waves
$\psi_\pm(\vec r,t)$.

Measurement is brought about by a spin-dependent interaction
\be
\hat H_{meas} = \hat H_0 ~+~ |+\rangle\langle+| ~ V_{meas}^+(\vec r)
  ~+~ |-\rangle\langle-|~ V_{meas}^-(\vec r),
\label{model}
\ee
where the measurement interactions $V_{meas}^{\pm}$ induce
spin-dependent dynamics of the c.o.m. wave function, tending to drive
the two components $\psi_{\pm}(\vec r,t)$ of the detector apart from
each other, whereas the spin-independent operator
\be
\hat H_0=-\frac{\hbar^2}{2M}\nabla^2 + V^0_{meas}(\vec r)
	+ V_{gr}(\vec r)
\label{h0}
\ee
contains -- besides the inactive apparatus terms -- the gravitational 
self-interaction of the detector, counteracting both the spreading 
of each partial wave component $\psi_{\pm}(\vec r,t)$ and their 
separation under the action of $V_{meas}^{\pm}$. As long as {\em both} 
effects remain in the range $r \ll R$, Eq.(\ref{gravpot}) remains 
valid, $\vec r~^*$ being now the common c.o.m of the split wave packets.

In view of the entanglement of the two partial wave packets of the 
apparatus to two orthogonal states of the microobject\cite{dec}, 
$\psi_+$ and $\psi_-$ can be considered as two classical objects
centered around $\vec r~^+$ and $\vec r~^-$ resp., and
their dynamics can be directly inferred from Ehrenfest's theorem. 
According to that theorem, accelerations are controlled by the 
quantum mechanical mean values of the forces corresponding to various
terms in Eq.~(\ref{model}). The averages should be calculated
in state (\ref{neumann}). In particular, the gravitational forces that
glue the partial wave packets together cancel out for the compound
object, in accordance with Newton's 3rd law:
\be
{\mathcal F}_{gr}=-M\omega_{gr}^2\bigl(|c_+|^2 (\vec r~^+-\vec r~^*) 
    ~+~|c_-|^2 (\vec r~^--\vec r~^*) \bigr)=0
\label{n3}
\ee
in view of the definition of the c.o.m. Therefore the acceleration of
the bound double wave packet as a whole is controlled by the quantum 
mechanical average of the non-gravitational forces:
\be
{\mathcal F}_{meas}=-\nabla
   \bigl(V_{meas}^0~+~|c_+|^2 V_{meas}^+~+~|c_-|^2 V_{meas}^-\bigr).
\label{ftot}
\ee
This important result says that in the classical domain the dynamics
of the measuring apparatus is governed by  modulus-squared quantum 
amplitudes acting as external classical fields, for unlimited time
scales. 

\subsection{Generation of randomness}

Since classical mechanical systems are capable to generate random 
behavior through a number of mechanisms, Eq. (\ref{ftot}) is certainly 
a valueable clue to the understanding of Born's law of probabilities. 
As a matter of fact, systems responding to external fields in a linear 
way are plentiful around us: any device of sufficiently strong 
ohmic--stochastic response is a quantum detector obeying
Born's rule. 

Our schematic models cannot be applied to real detectors without
considering their two-stage, photographic mode of operation: fast
passage of a particle acts as a kind of exposure, leaving a latent
trace in the form of ionization or excitation, fully within the range 
of linear quantum mechanics; that trace is subsequently developed into 
a macroscopic signal at a comfortable pace, open to gravitational
control. Either stage can exhibit some mechanism of amplification, 
often based on metastability \cite{vankam}. In a photomultiplier or 
avalanche photodiode \cite{sze} the first stage is the avalanche 
formation itself, the second the rise of a voltage pulse in the 
external circuit; in a cloud (or bubble) chamber the first is the 
formation of a trace of ionization \cite{mott}, the second the 
condensation of liquid droplets (or bubble formation in a superheated
liquid). The exposure and development stages are obviously 
associated to different degrees of freedom. 

In what follows, we leave those complications and turn back to a
highly schematic model for a metastable detector, expected to give a 
signal for, say, $\sigma_z=+1$, triggered by the Ehrenfest-averaged 
measurement force ${\mathcal F}=-|c_+|^2~\nabla~ V_{meas}^+$. As a 
specific feature of our model, capable to exhibit eventual deviations 
from Born's rule, we assume that our detector is biased to threshold. 

Accordingly, we consider a potential well open along some axis $\xi$ 
(``measurement coordinate'') through a saddle point of the potential:
\be
V_{meas}^0(\vec r)=a\xi-b\xi^3+c(r_\perp^2)
\label{saddle}
\ee
where $r_\perp$ is the orthogonal distance from axis $\xi$. Classically,
for a given energy of the detector, this is a chaotic billiard; closed
if its energy is below the height of the saddle, open if not. In the 
threshold case, in absence of interaction with the microobject the 
energy should be at the saddle. As interaction starts, the mean 
measurement force ${\mathcal F}$ would lower the saddle point to 
first order by  $(a/3b)^{1/2}{\mathcal F}$, and admit barrier crossing 
inside the circle 
\be
r_\perp\leq\Bigl(\frac{a}{3bc^2}\Bigr)^{1/4}{\mathcal F}^{1/2}, 
\label{circle}
\ee
at a speed bounded by 
\be
v_{max}=\Bigl(\frac{4a}{3bm^2}\Bigr)^{1/4}{\mathcal F}^{1/2}. 
\label{vmax}
\ee
Variations of density and speed over the area of the circle cancel out. 
Finally, the rate of escape from the potential well is obtained as
\be
\Gamma\propto(\pi r_\perp^2)\cdot v_\parallel.
\label{rate}
\ee 
The first factor being proportional to ${\mathcal F}$, the second to 
${\mathcal F}^{1/2}$, the escape rate and thereby the probability of
detection becomes proportional to the 3/2-th power of $|c_+|^2$, 
which - although qualitatively in the usual way - does not agree with 
Born's rule. 

By good luck, real detectors do not operate in that way: they are
biased high above threshold, and their signals are cut by some
electronic discriminator. That linearizes the $|c_+|^2$ dependence and
Born's rule is recovered. 

Nevertheless, there is a  lesson to be drawn from the above
derivation, notwithstanding the oversimplified character of the model: 
Born's rule of strict proportionality between detection probabilities 
and modulus-squared amplitudes may depend on details of the underlying 
nonlinear dynamics. Particle detection at very low quantum amplitudes 
(produced by very asymmetric beamsplitters e.g. along the line of Ref. 
\onlinecite{gran}), accompanied by an exploration of non-optimized bias
and discrimination levels, possibly extended to EPR-like tests
\cite{epr} to see if remote detector correlations surpass Born's rule 
in robustness, is a promising target to detect eventual deviations. 

\section{NUMERICAL ESTIMATES}

Now we turn back to the question of whether gravity is strong enough 
to enforce classicality as outlined in the Introduction. According to
Eq. (\ref{class}),  that depends on the accelerations gravitational
confinement can hold out.

As a first test, we choose a spherical body of mass density 
$\varrho_0 \simeq 10^4~Kg/m^3$, typical for condensed matter, down to 
molecules. Then, an acceleration of $1 nm/(1 s)^2$, easy to detect
e.g. in an Atomic Force Microscope, gives a size limit $R\gg 1 mm$ and 
a corresponding mass $M\gg 10^{-2} g$ as a condition for
gravity-stabilized classicality, which -- although somewhat too large 
-- seems to be on the borderline of reasonability.

That was the best case. The most widespread case of a real quantum 
measurement process, all-electronic detection of particles, is the worse. 
The most massive accelerating component is the electron liquid of
electron density $10^{30}/m^3$ and corresponding mass density 
$\simeq 1 Kg/m^3$ in the detector - to - voltmeter leads, and 
it is far from satisfying criterion (\ref{class}). As a matter of
fact, in Geiger mode detection, that electron liquid transfers 
the charge of an avalanche of $10^7$ electrons to the voltmeter
through a lead of cross-section $10^{-9}m^2$  within $10^{-8}~s$, 
which corresponds to $a_{max}=10^2~m/s^2$, resulting in astronomical 
sizes required. The geometrical factors referring to non-spherical
shapes, presented in the Appendix, offer no sizeable modification.

One important point can save the scheme from being rejected on the
basis of the above numerical estimates: some recent theoretical 
developments indicate that in the submillimeter range new 
gravitation-like forces may arise, stronger than Newtonian gravity 
by factors up to $10^6$. The effect would be due to large extra 
dimensions \cite{extra}, with an eventual further contribution from 
low-energy supersymmetry breaking \cite{short}. Experiments have 
found no evidence so far \cite{oldex} but continue to explore 
shorter distances \cite{newex}.

One last point merits discussion here. As mentioned in the
Introduction, the very existence of the non-linear self-interaction, 
measured e.g. by the constant term $V^0_{grav}$ appearing in 
Eq.~(\ref{gravpot}), is in principle open to interferometric detection 
through the phase difference between two geometries differing in the 
respective positions of an absorber \cite{shinl}. For a given density 
$\varrho_0$ the effect scales with $M^{5/3};$ using the above figures 
and a typical time-of-flight of $10^{-2} s,$ as a necessary condition 
for a phase shift of order unity one obtains $M\approx 10^{-14} Kg,$ 
which is by ten orders of magnitude heavier than present-day limits of 
molecule interferometry \cite{zeil}. A more detailed analysis of 
interferometric possibilities \cite{hasszan} does not change that 
conclusion. However, if -- as mentioned above -- gravity proves much 
stronger on the nanoscale, that may reopen the issue of
interferometric detection.

\section{THE NONLOCALITY ISSUE}

Now we turn to the questions of entanglement-based correlations. To
begin with the easy part of it, according to a subtle generic
objection to nonlinear versions of quantum mechanics \cite{bbmw}, they
are liable to give rise to superluminal signaling, either between remote
entangled subsystems or between Everettean branches of a superposition
\cite{gispol}, thereby breaking the fragile ``peaceful coexistence''
between quantum mechanics and special relativity \cite{shipc}. The
present scheme is free from that kind of objection: on the one hand,
as discussed by Polchinski\cite{gispol}, EPR signaling between
entangled partners requires collapse of the wave function which is
absent here; on the other hand, Everettean branches for macro-objects
are now not separated macroscopically, therefore signaling between
them remains harmless \cite{signal}.

There is, however, a major difficulty of the Newton-Schr{\"o}dinger scheme:
its inability to explain the dynamical origin for the robust
manifestations of entanglement through anticorrelations between two or
more detectors probing a single quantum system \cite{epr,ei}.

Orthodox Copenhagen quantum mechanics is free from that difficulty, at
the expense of disclaiming to look for a dynamical origin of
statistical laws. We illustrate that on the simplest case of
1-particle 2-detector anticorrelations, i.e. the fact that out of
two detectors awaiting a single particle, only one would give a signal
\cite{ei}.  In a Stern-Gerlach arrangement with two spatially
separated partial waves of the microobject, covered by their
respective detectors labelled 1 and 2, unitary evolution
would result in the entangled quantum state
\be
|\Psi\bigr\rangle
  \Rightarrow
         c_+|+\rangle\psi_+^{(1)}\psi_-^{(2)}~
        +c_-|-\rangle\psi_-^{(1)}\psi_+^{(2)};
\label{twodetect}
\ee
measurement would take place by the quantum state collapsing into one
of the above terms. Either of the choices implements full
anticorrelation of the detectors.

The main difference with the present approach is that whereas in
orthodox quantum mechanics $\psi_+{(i)}$ and $\psi_-{(i)}$ ($i=1$
or $2$) are macroscopically distinct, our approach considers them
as being lumped together by gravity, which is, however, acting
independently on the two detectors. Accordingly, the two
non-collapsing Everettean branches of the superposition are
macroscopically indistinguishable, moving as a single point of the
configuration space spanned by some measurement coordinates $\xi_1$
and $\xi_2$ of the two detectors which, however, take their classical
random choices independently. Apparently, there is nothing in the
formalism to exclude that both detectors should choose to be active,
or both quiescent. The same refers to the much-tested  EPR case 
\cite{epr}: if detectors take independent random choices,
there is nothing in the scheme to force anticorrelations.

\begin{figure}
\includegraphics{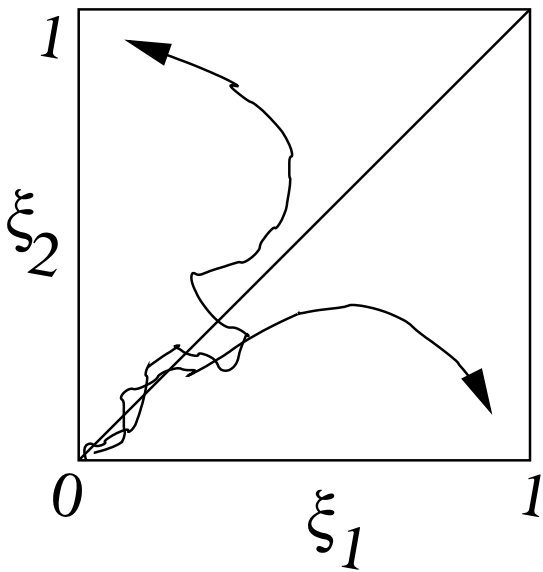}
\caption{\label{meas1.eps} Anticorrelation of two detectors awaiting 
one particle: on the two-dimensional flow diagram of the respective 
measurement coordinates $\xi_1$ and $\xi_2$ of the detectors (0 if 
quiescent, 1 if active), there seems to be some force, destabilizing  
the diagonal and driving the system towards one of the possible end 
points with one detector active, one quiescent.} 
\end{figure}

However, detector anticorrelation is a strong effect. If, like
randomness itself, it has a dynamical origin, that origin must be a
strong force, destabilizing  the ``diagonal'' in the two-detector
configuration space \cite{berry} and driving the system towards one 
of the possible end points with one detector active, one quiescent 
(see FIG.~\ref{meas1.eps}). To make it more embarrassing, this is 
not a genuine configuration space, detector anticorrelations being
completely detached from the detectors' location in three-dimensional
space. This is fully analogous to the situation in the stochastic 
reduction scheme\cite{grwd}: if there is a dynamical process behind 
the correlations, whether it is described by an external random force 
\cite{gisepr} or by a nonrandom one as suggested here, it must be 
related to the whole entangled state vector; so far, with no better 
representation in sight than Hilbert-space itself. Where to look in the 
theory for such a force, remains open to speculation for the moment. 
My guess and hope, already hinted at in the introduction to this
Letter, is a possible modification of the Newton-Schr{\"o}dinger scheme 
\cite{di} in such a way that the source of gravity  should include 
quantum entanglement. The possibly related open issues of dark matter 
and Einstein's cosmological term may add to the identikit of the desired 
modification. 

\section{CONCLUSIONS}

In this paper a thorough critical study of the Newton-Schr{\"o}dinger 
approach to the quantum measurement problem has been presented, under
the ambitious scope of supporting a no-split no-collapse scenario,
with emergent randomness. The qualitative picture formed has extremely 
promising features. The scheme offers a transparent characterization 
of classicality, which is self-sustaining; it eliminates the
paradoxical features of the Everett approach since macro-objects do
not suffer macro-separation; finally, it offers a dynamical route to 
emergent quantum randomness. 

However, there are serious doubts preventing immediate application
of the scheme. First of all, gravity, as we know it from macroscopic 
measurements, is too weak for the purpose. Therefore its eventual 
short-range enhancement, theoretically proposed but not yet
confirmed by straightforward experiments, may be crucial for
classicality. In turn, if the present scheme gets further
support, that can serve as an indirect indication for short-range
gravity enhancement.

More far-reaching is the issue of the unexplained origin of 
entanglement-based detector anticorrelations, which calls for a
refined characterization of the source of gravity, with nonlocal 
forces related to quantum entanglement in a way to be found.

In view of the discussion in the previous section, that is a
formidable task. In the same time, until explicitly falsified, it 
is a logically open - even if very narrow - way out of the dual 
description of the quantum-classical world.

\section*{ACKNOWLEDGMENTS}

\noindent I thank Lajos Di{\'o}si for giving me innumerable explanations
about various approaches, as well as Abner Shimony and Nicolas Gisin
for important criticism on various preliminary versions. I am indebted
to Anton Zeilinger for supplying reference \cite{gran}. Discussions
with my MSc student Istv\'an Varga-Haszonits helped clarifying some
points. This work has been partially supported by the Hungarian
Research Foundation (grants OTKA T 029544 and T 034832).

\appendix
\section{}

\noindent The values of the parameters appearing in
Eq. (\ref{gravpot}) are given in the for a spherical body of radius
$R$. Here, motivated by the case of an electron liquid in a metallic
lead connecting a detector to a voltmeter, we give them for the case 
of a rectangular slab of cross-section $a\times b$
(diameter $d$) and length $L\gg a,b$:
\be
V^0_{gr} = -2\ln(L/d)GM^2/L \nonumber
\ee
and
\be
\omega_{gr}^2=G\varrho_0\frac{d}{L}f\bigl(\frac{a}{b}\bigr),
\nonumber
\ee
where
\bea
f(x)=\frac{2}{x+1/x}\Bigl(2\sqrt{1+x^{-2}}\ln(x+\sqrt{1+x^2})\nonumber\\
+2\sqrt{1+x^2}\ln(x^{-1}+\sqrt{1+x^{-2}})\nonumber\\
-\frac{2}{3}(2+x^2+x^{-2}-x\sqrt{1+x^2}-x^{-1}\sqrt{1+x^{-2}})\Bigr)
\eea
In particular, $f(x)=4.205$ for $x=1$ and $(4/x)\ln x$ for $x\gg 1$.

\end{document}